\documentclass[twocolumn,showpacs,prb,floatfix,superscriptaddress,
citeautoscript]{revtex4}
\usepackage{graphicx}
\usepackage{color}
\usepackage{amsmath}

\usepackage{booktabs}

\begin{document}

\title{GraphAne: From Synthesis to Applications}

\author{H. Sahin}
\email{hasan.sahin@uantwerpen.be}
\affiliation{Department of Physics, University of Antwerp, 2610
Antwerp, Belgium}

\author{O. Leenaerts}
\email{ortwin.leenaerts@uantwerpen.be}
\affiliation{Department of Physics, University of Antwerp, 2610
Antwerp, Belgium}

\author{S. K. Singh}
\email{sandeepkumar.singh@uantwerpen.be}
\affiliation{Department of Physics, University of Antwerp, 2610
Antwerp, Belgium}

\author{F. M. Peeters}
\email{francois.peeters@uantwerpen.be}
\affiliation{Department of Physics, University of Antwerp, 2610
Antwerp, Belgium}

\date{\today}


\begin{abstract}

Atomically thin crystals have recently been the focus of attention in
particular after the synthesis of graphene, a monolayer hexagonal crystal
structure of carbon. In this novel material class the chemically derived
graphenes have attracted tremendous interest. It was shown that although
bulk graphite is a chemically inert material, the surface of single
layer graphene is rather reactive against individual atoms. So far, synthesis
of several graphene derivatives have been reported such as hydrogenated
graphene "graphane" (CH), fluorographene (CF) and chlorographene (CCl). 
Moreover, the stability of bromine and iodine covered graphene were predicted 
using computational tools. Among these derivatives, easy synthesis, insulating 
electronic behavior and reversibly tunable crystal structure of graphane make 
this material special for future ultra-thin device applications. This overview, 
surveys structural, electronic, magnetic, vibrational and mechanical properties of graphane. We also present a detailed overview of research efforts devoted to the computational  modeling of graphane and its derivatives. Furthermore recent progress in  synthesis techniques and possible applications of graphane are reviewed as well.

\end{abstract}

\maketitle

\section{Introduction}

During the last decade, the field of materials science has been expanded with
the advent of few-atom-thick two-dimensional materials. Previously, the
majority of research was devoted to three-dimensional bulk crystals
because these are easier to handle in experiment and in theoretical
simulations. The ever increasing trend to miniaturization in the last decades
has been accompanied with the production of low-dimensional nanomaterials in all
but two dimensions. The accepted theoretical impossibility of such 2D materials
has probably been the largest obstacle for their fabrication. 2D systems, such
as electron gases, have been known for quite some time, but they usually appear
at the surface or interface of 3D materials. In material science, the
dimensionality of a material is not a strictly defined quantity, but it is
common to refer to few-atom-thick layers as 2D. In that sense, 2D materials have
only come to the center of attention with the isolation and characterization of
graphene in 2004.\cite{nov04,nov05}

The production of graphene was originally achieved with the now famous
scotch-tape method. Starting from a highly-ordered pyrolytic graphite sample,
few-layer graphene samples are isolated by repeated peeling with an adhesive
tape, and subsequent deposition on a substrate (usually silicon dioxide).
Nowadays, many different techniques have been developed to produce graphene from
other (3D) crystals such as silicon carbide or using bottom-up approaches.

Graphene is a unique material in many ways. It is the thinnest imaginable
crystal but, at the same time, it is the strongest\cite{lee08}, stiffest, and
best conducting material\cite{bal08} known. Furthermore, it has an intriguing
electronic spectrum
that mimics that of massless Dirac particles, leading to many fascinating
phenomena such as an anomalous integer quantum Hall effect and Klein tunneling.
The absence of a band gap in the electronic spectrum of graphene can also be
considered as an annoying feature that prohibits the direct implementation of
graphene in electronics. Therefore, a large part of the research on graphene
has been devoted to the creation of an electronic band gap. This can be
achieved in many ways: (i) through a reduction of the dimensionality by cutting
graphene into ribbons or flakes, (ii) by periodic potentials or substrates that
break the sublattice symmetry, or (iii) by the destruction of the $\pi$-bond
network through functionalization.

\begin{table*}[htbp] \centering
\caption{The relative stability of various graphane isomers with respect to the
\textit{chair} isomer as found in the literature. The total energies,
calculated for per unitcell, are given in eV.\label{tab-configurations}}
\begin{tabular}{lccccccccc}
\hline\hline
               & Sluiter  & Sofo    & Samarakoon   & Leenaerts & Artyukhov &
Wen
& Samarakoon  & Bhattacharya & He \\
               & Ref.~\ \onlinecite{slu03} & Ref.~\ \onlinecite{sof07}  &
Ref.~\
\onlinecite{sam09} & Ref.~\ \onlinecite{lee10} & Ref.~\ \onlinecite{art10} &
Ref.~\
\onlinecite{wen11} & Ref.~\ \onlinecite{sam11} & Ref.~\ \onlinecite{bha11} &
Ref.~\
\onlinecite{he12}\\
\hline
  \textit{chair}			&	0.000 & 0.00 & 0.00 & 0.000 &
0.000	&  0.00  & 0.00	& 0.00	& 0.000 \\
  \textit{boat}				&	0.103 & 0.12 & 0.17 & 0.103 &
0.103 &  0.10  & 0.09	& 0.10	& 0.102 \\
	\textit{stirrup } 	& 0.056 &  -   &  -   & 0.053 & 0.055	&  0.05

& 0.05	& 0.04	& 0.056 \\
  \textit{armchair}		&	  -   &  -   &  -   & 0.128 &   -  	
&  0.13  &  - 	&  -  	& 0.134 \\
  \textit{TB-chair}		&	  -   &  -   & 0.25 &   -   &   - 	
&  0.19  &  -	  &  -  	&   -   \\
  \textit{twist-boat} &   -   &  -   &  -   &   -   &   - 	&   -    & 0.16	
&  -  	& 0.150 \\
  \textit{tricycle}   &   -   &  -   &  -   &   -   &   - 	&   -    &  - 	
&  -  	& 0.024 \\
\hline\hline
\end{tabular}
\end{table*}

Functionalization of graphene can be achieved in different ways. It is possible
to substitute some C atoms in the graphene layer by foreign atoms such as B and
N. In this way charge doping can be obtained and a small band gap can be
opened. Another way of functionalization is through the adsorption of atoms or
molecules. Depending on the adsorption strength, one can distinguish between
physisorption and chemisorption. Physisorption leaves the graphene structure
virtually unaltered, but can induce some subtle electronic changes such as
charge doping. Chemisorption, on the other hand, is accompanied by strong
covalent bonds between the graphene C atoms and the adsorbate. These covalent
bonds disrupt the aromatic network in graphene and causes strong structural
changes. The adsorbates that can cause these kind of changes are usually
radicals such as hydroxyl groups or atomic species. When the coverage of
chemisorbed adsorbates is dense enough, the properties of graphene can change
drastically, as in the case of oxygenation.\cite{jun08,jia09} A (reversible)
metal to insulator transition is predicted upon covalent
functionalization.\cite{eza13}

Depending on the type of adsorbate the functionalized graphene can exhibit
crystalline or amorphous character. The former is expected for hydrogenation
and fluorination while the latter has been observed for oxygenated graphene
(graphene oxide). The
distinction between crystalline and amorphous is not always clear and can depend
on the level of coverage. Furthermore it is expected that island formation
occurs\cite{bal10}, which makes the distinction even more troublesome.

The chemical functionalization of graphitic materials has a longer experimental
and theoretical history that dates back from before the isolation of graphene.
Fluorinated graphite and fluorinated graphene have been thoroughly
studied\cite{rud47,ebe74,wat88,cha93} and much of
the obtained knowledge can be transferred to the case of hydrogenated graphene.

\section{Properties of Graphane}

Graphane is the name given to a class of hydrogenated graphene structures that
are close to full coverage. Ideally, every carbon atom of the graphene layer is
covalently bonded to a hydrogen atom. These C-H bonds can only exist if the C
atoms in graphene change their hybridization from $sp^2$ to $sp^3$ and, as a
consequence, the graphene layer becomes buckled. There is no
unique way to couple every C to a H atom because the H atoms can attach to the
graphene layer from above or below. In fact, there is also no \textit{a priori}
reason why this should be done in an ordered fashion, but to keep calculations
and interpretations simple, one usually supposes the graphane structure to be
crystalline. It is assumed that the H atoms are chemisorbed according to some
simple pattern and during the last decade, many different patterns have been
suggested. These are usually referred to as the different conformations,
configurations, or (stereo) isomers of graphane. Note that the different CH
structures
are actually not conformers because they are not related by mere bond rotations.
In this section, we take a closer look at the structure and physical properties
of the different graphane isomers. We review their electronic, vibrational,
and optical properties and comment on their relative stability.

\begin{figure}[htbp]
  \centering
  \includegraphics[width=8.5cm]{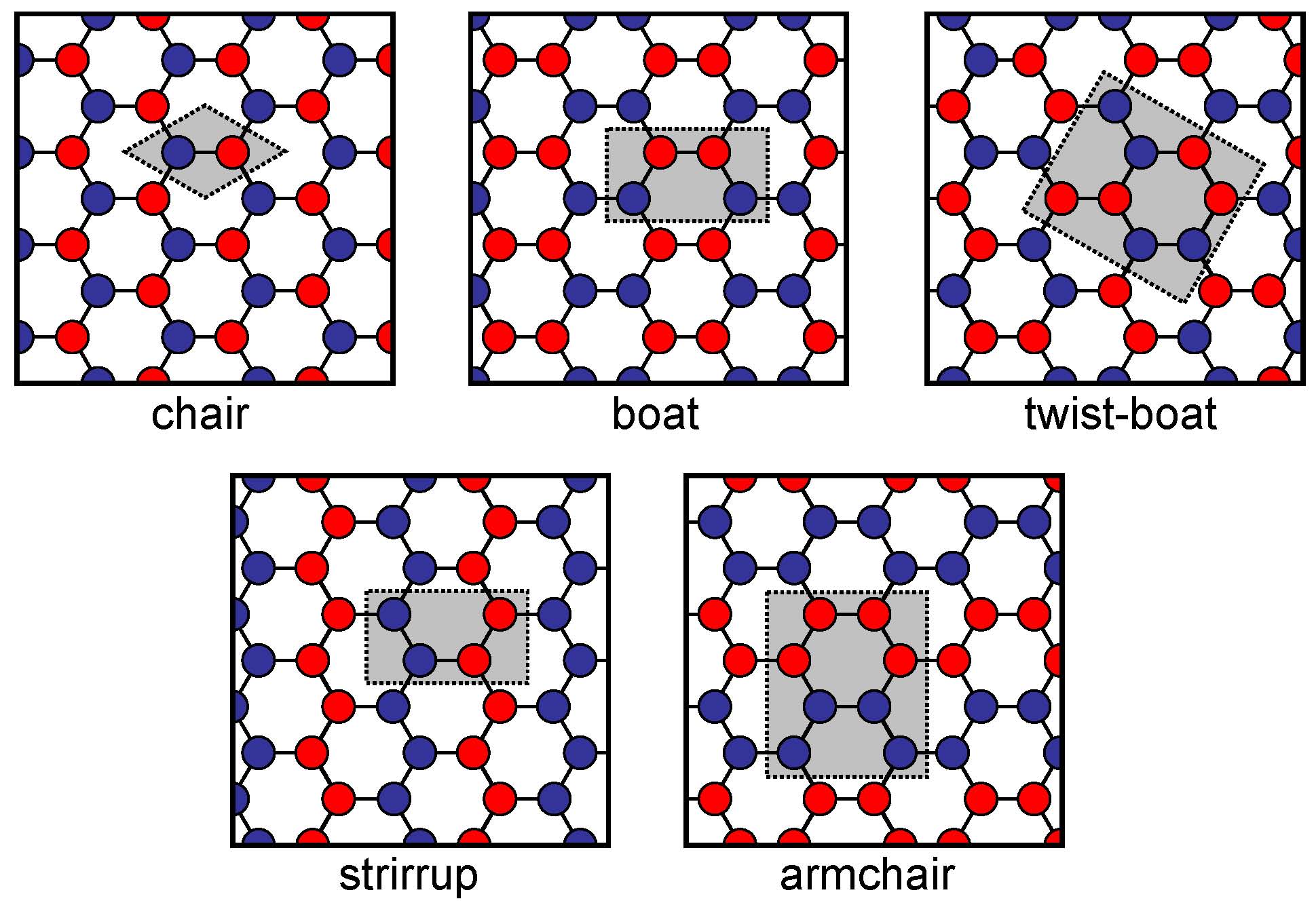}
  \caption{Five isomers of graphane in which every C atom is equivalent. Blue
and red colors indicate H adsorption, respectively, above and below the
graphene layer.\label{fig_iso}}
\end{figure}

\subsection{Atomic Structure}
\begin{figure}[]
  \centering
  \includegraphics[width=8.5cm]{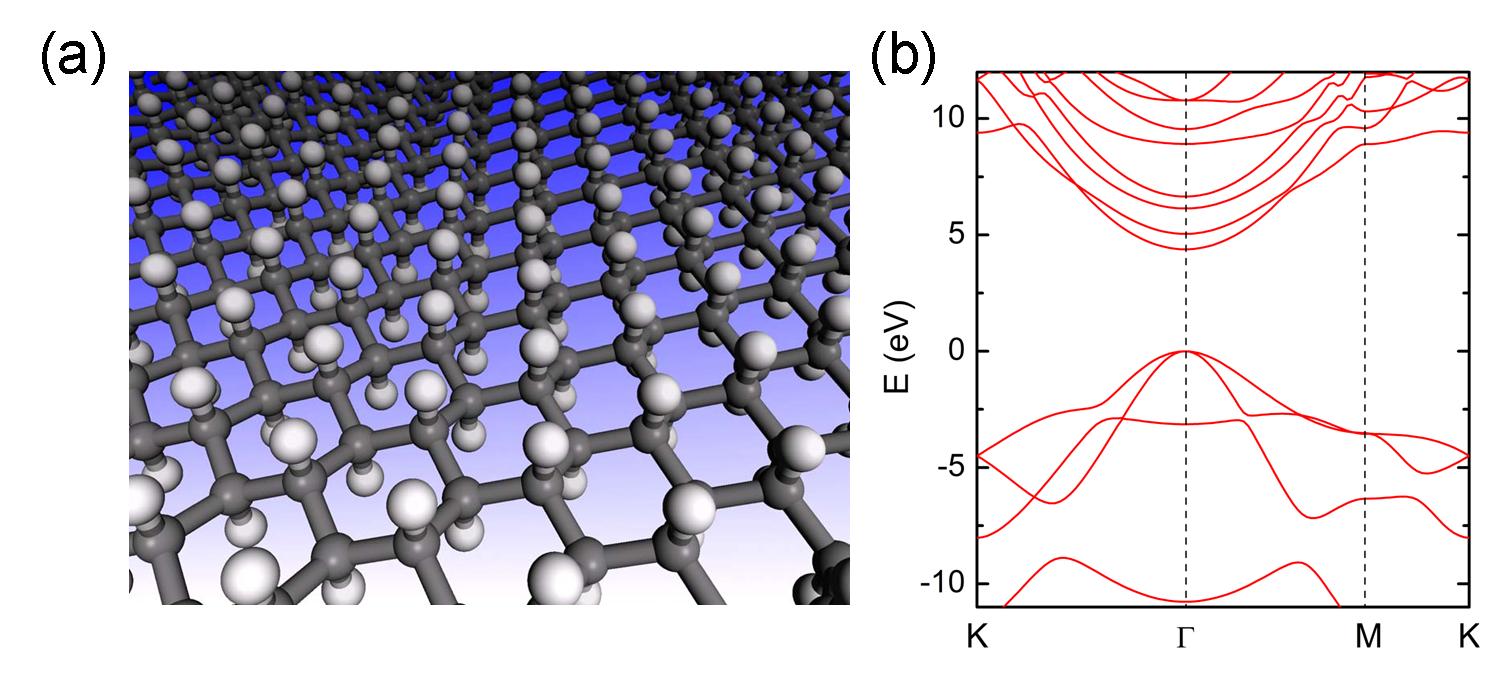}
  \caption{(a) Crystal structure of graphane in \textit{chair} configuration
  from Ref. ~\onlinecite{sof07}. (b) Band structure of graphane from
  Ref. ~\onlinecite{lee10}.\label{fig_BS}}
\end{figure}

In Fig.\ \ref{fig_iso}, we show the schematic structure of the three best-known
and most stable graphane isomers, namely the \textit{chair}, \textit{stirrup},
and \textit{boat} configuration, together with two other isomers. The former
three structures have already been proposed in 2003, i.e.\ before the isolation
of graphene\cite{nov04}, by Sluiter and Kawazoe\cite{slu03}. However, it took
until 2007, when Sofo et al. rediscovered\cite{sof07} the
\textit{chair} and \textit{boat} configurations, that graphane acquired
wide-spread attention from the graphene community. Since then, more structures
have been proposed such as the \textit{twist-boat}\cite{sam11},
\textit{twist-boat-chair}\cite{sam09}, \textit{armchair}\cite{lee10},
\textit{tricycle}\cite{he12}, and many others, but they are all considerably
less stable than the known configurations (except for the \textit{tricycle
}configuration which is actually a combination of \textit{chair} and
\textit{stirrup}) and are therefore of minor importance. In table
\ref{tab-configurations}, we compare the calculated stability of different
graphane configurations that can be found in the literature. All computational
studies agree that the \textit{chair} configuration is the most stable one. In
this configuration the H atoms are alternately adsorbed above and below the
graphene sheet so that all the C atoms of one sublattice move up while those of
the other sublattice move down (see Fig.\ \ref{fig_BS} (a)). \textit{} Although
less common in the literature, the \textit{stirrup} configuration (also called
\textit{washboard}, or \textit{zigzag} configuration), is more stable than the
so-called \textit{boat} configuration. The \textit{stirrup} isomer, schematically 
shown at the bottom left of Fig.\ \ref{fig_iso}, consist of alternating zigzag chains
with H atoms pointing up and down. The reason that this graphane
isomer is less well-known is probably the fact that the important paper of Sofo
\textit{et al.}\cite{sof07} considered the \textit{boat} configuration, but not the
\textit{stirrup} one.

In discussing the dependence of the properties on the type of conformation, we
take the four isomers discussed by Leenaerts et al.\cite{lee10} as
characteristic examples. Wen et al. showed that these isomers are
dynamically more stable than benzene molecules.\cite{wen11} The data
for the structure of these graphane isomers are given in Table
\ref{tab-struct}.

\begin{table}[h] \centering
\caption{Structure parameters for the different graphane isomers. Distances are
given in $\text{\AA}$ and angles in degrees.\label{tab-struct} The distance
between neighboring C atoms, $d_{\textsc{cc}}$, and the angles,
$\theta_{\textsc{ccx}}$, are averaged over the supercell.}
\begin{tabular}{lcccc}
\hline\hline
                                      & chair & boat  & strirrup& armchair\\
\hline
  $a_x/\sqrt{3}$								
	
		& 2.539 & 2.480 & 2.203 & 2.483   \\
  $a_y/n_y$									
	
				& 2.539 & 2.520 & 2.540 & 2.270   \\
	$d_{\textsc{ch}}$  	 						
	
	& 1.104 & 1.099 & 1.099 & 1.096   \\
  $\overline{d}_{\textsc{cc}}$				& 1.536 & 1.543 & 1.539
& 1.546  	\\
  $\overline{\theta}_{\textsc{cch}}$	& 107.4 & 107.0	& 106.8	& 106.7		
\\
  $\overline{\theta}_{\textsc{ccc}}$ 	& 111.5	& 111.8	& 112.0	& 112.1		
\\
\hline\hline
\end{tabular}
\end{table}

The length of the C-H bonds is about 1.10 $\text{\AA}$ which is typical for
such bonds as found in organic chemistry. The average length of the C-C bonds
is close to the ideal single C-C bonds in diamond. Since there is no steric
strain in the chair configuration, because all bonds and angles are completely
free to relax, the bond lengths and angles can be considered ideal in that
case. Deviations from this ideal situation are found in the other
configurations which indicate the presence of local strain. Some of the isomers
contain substantial structural anisotropies which are translated into their
electronic and mechanical properties (see below).

\subsection{Electronic Structure}
Graphane is a semiconducting material with a substantial direct electronic
band gap (see Fig.~\ref{fig_BS}(b)). The size of the gap depends on the method
of calculation: within DFT the gap is about 3.5 eV for LDA and GGA, and 4.4 eV
with a hybrid functional (HSE06)~\cite{hey06}. The hybrid functional (HSE)~\cite{hey03,hey06} 
overcomes some deficiencies of the conventional approximations for the exchange-correlation 
(xc) functional within DFT, i.e. LDA or GGA. The later include an unphysical 
interaction of the electron with itself, the so-called self- interaction, yielding 
a systematic underbinding of strongly localized states. HSE partly corrects for this spurious 
self-interaction by intermixing a fraction of Hartree-Fock exchange. This leads to
an improvement of the structural (equilibrium lattice constants, bulk moduli), thermochemical 
(cohesive energies, heats of formation), and electronic properties (band gaps) 
of semiconductors and insulators.~\cite{pai06} Still higher accuracy on the electronic 
band gap can be achieved by including many-body interactions (GW approximation). 
Such many-body interactions increase the electronic band gap of graphane further 
to about 5.2 eV\cite{leb09,lee10}. 

\begin{figure}[]
  \centering
  \includegraphics[width=7.5cm]{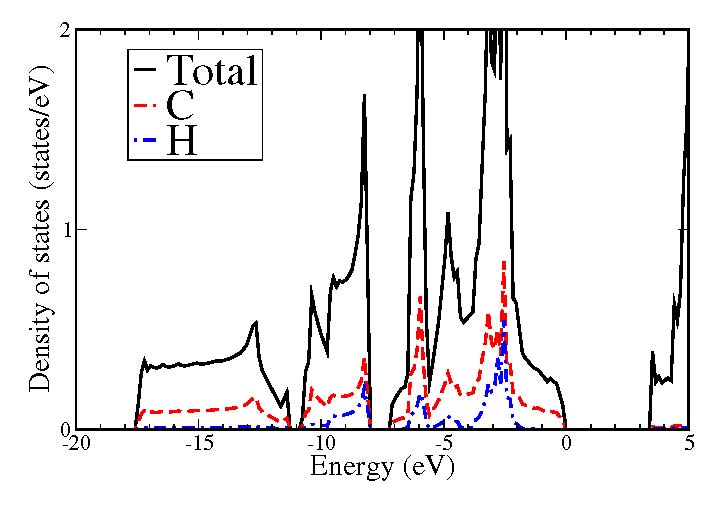}
  \caption{The total and projected density of states of graphane. From Ref. \
\onlinecite{leb09}. \label{fig_PDOS}}
\end{figure}

The band gap is almost independent of the
configuration and varies with less than 5\%.\cite{lee10} The total and projected
density of states (DOS) of graphane in the \textit{chair} configuration is shown
in Fig.~ \ref{fig_PDOS}.  The electronic gap is clearly recognizable in the
total DOS, but the projected DOS shows some peculiar characteristics. The
electronic states below the gap are localized on the C atoms but the
contribution of C and H atoms to the conduction band states is negligible.

\begin{figure}[htbp]
  \centering
  \includegraphics[width=8.5cm]{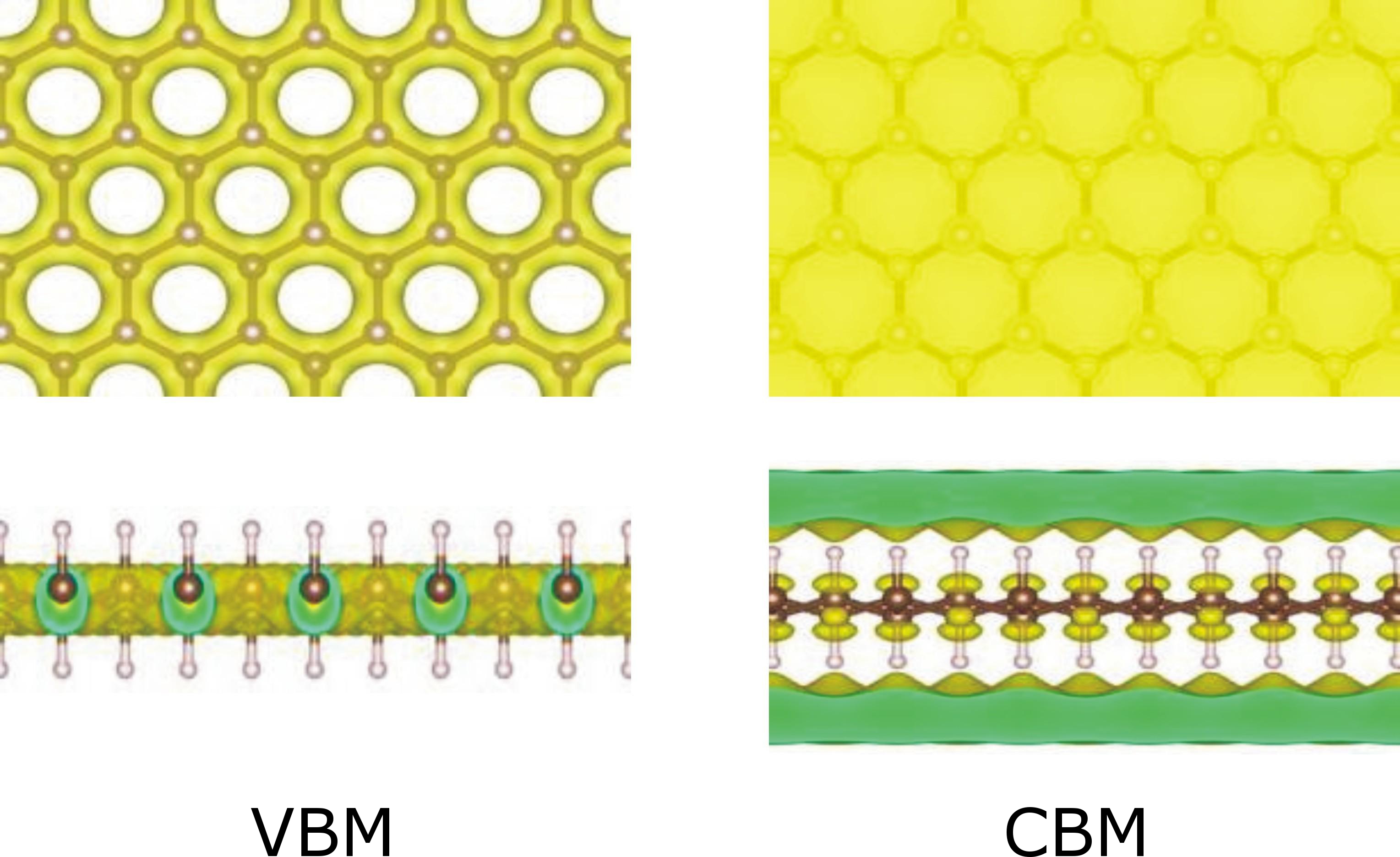}
  \caption{Top and side views of the VBM (left) and CBM (right) states of
graphane. From Ref. \ \onlinecite{moz13}.\label{fig_WF}}
\end{figure}

This is even more clear when considering the density of the single-particle
states at the valence band maximum (VBM) and conduction band minimum (CBM), as
shown if Fig.\ \ref{fig_WF}. The VBM and CBM have very different character and
symmetry. The valence band states consist of C $p_x$ and $p_y$ orbitals, but
the conduction band states exhibit plane-wave character. The CBM state behaves
as a delocalized electron that is loosely bound above the H atoms.\cite{moz13}

\subsection{Optical properties}

\begin{figure}[htbp]
  \centering
      \includegraphics[height=1.7 in]{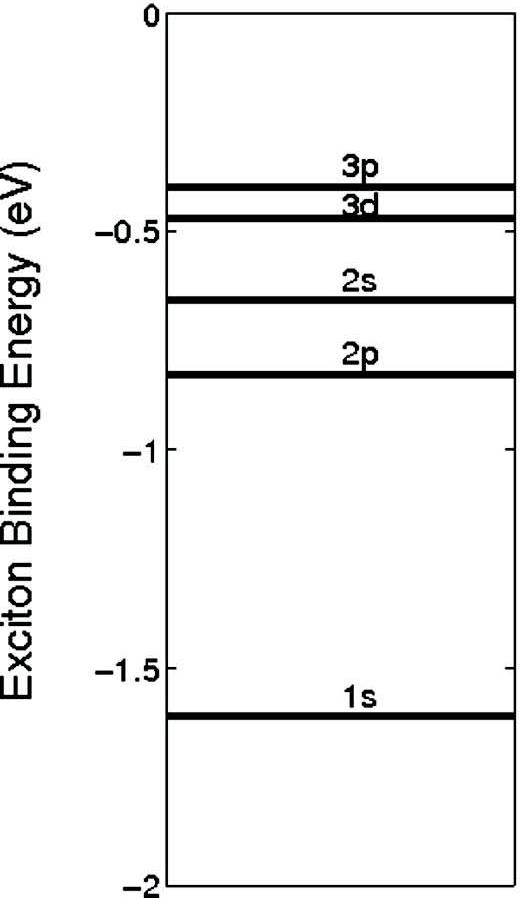}
      \includegraphics[height=1.7 in]{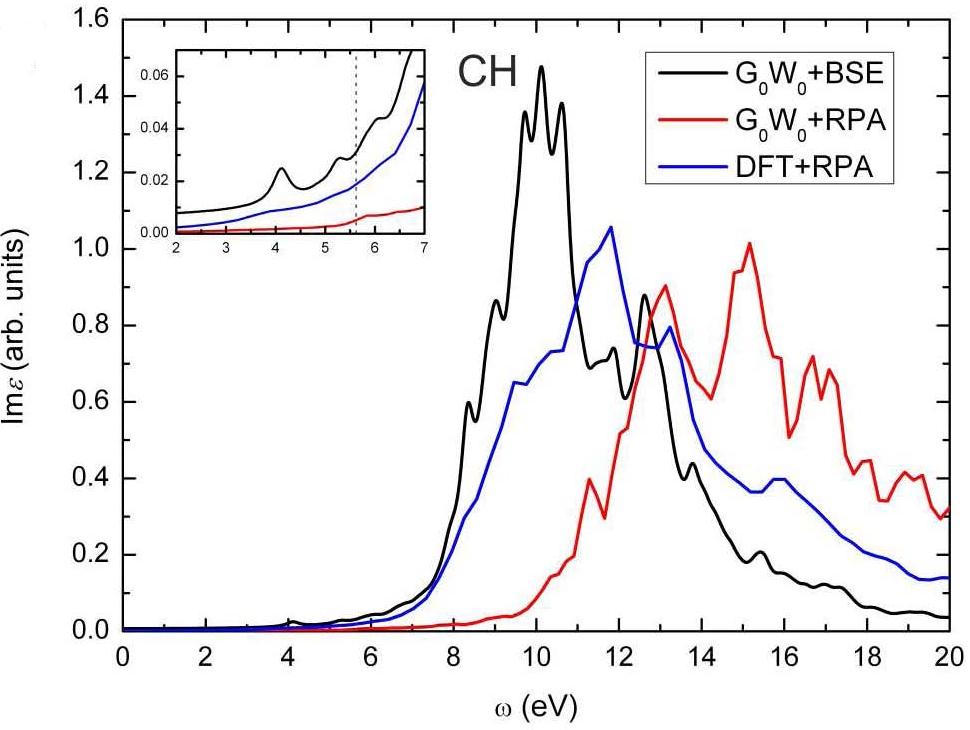}
  \caption{(Color online) (left) The excitonic spectrum of bound states in
graphane from Ref.\ \onlinecite{hua13}. (right) The imaginary part of the
dielectric function of graphane calculated within DFT (DFT+RPA), with
\textit{e-e} interactions (G$_0$W$_0$+RPA), and with \textit{e-e} and
\textit{\textit{e-h}} interaction (G$_0$W$_0$+BSE) from Ref. \
\onlinecite{kar13}.\label{fig-optical}}
\end{figure}

The two-dimensional nature of graphane has a significant impact on its optical
properties and reduces the optical band gap significantly. The reduced
screening in two-dimensional materials leads to large electron-hole
interaction.\cite{cud10} As a consequence, there are bound exciton states
considerably below the conduction band minimum. For graphane the exciton
binding energy has been calculated to be 1.6 eV\cite{hua13} which produces an
optical gap of about 3.8 eV.\cite{cud10,kar13,res14} A more complete exciton
spectrum is shown in Fig.\ \ref{fig-optical}(left). The increase of the electronic
band gap due to electron-electron interactions, is almost completely canceled by
the electron-hole interaction. This is nicely illustrated in Fig.\
\ref{fig-optical}(right) which shows the absorption spectra (imaginary part of the
dielectric function) at 3 different levels of approximation: (i) without
\textit{e-e} and \textit{e-h} interaction, (ii) with \textit{e-e} but no
\textit{e-h} interaction, and (iii) with \textit{e-e} and \textit{e-h}
interaction. Single-particle calculations (DFT) lead to similar band gaps as
calculations with \textit{e-e} and \textit{e-h} interaction (G$_0$W$_0$+BSE).

In experiments, one usually estimates the band gap from optical absorption
measurements. This kind of measurements actually measures the optical band gap
and not the electronic band gap. Therefore, one should take care to compare
experimental results with the appropriate theoretical calculations.

\subsection{Magnetic properties}
Fully hydrogenated graphene contains only $sp^3$ hybridized C atoms with 4
bonding partners (3 C and 1 H). Therefore, the ideal graphane isomers are not
expected to show any interesting magnetic properties. Magnetism only occurs at
defect sites (missing H atoms\cite{sah09,ber10}) and in partially hydrogenated
graphene. We discuss this in more detail below. Substitutional defects such as
B atoms, can also lead to a magnetic response in graphane.\cite{wan11}

Another important effect of hydrogenation of graphene is the enormous increase
in spin-orbit coupling.\cite{cas09,gmi13} The change in hybridization from
$sp^2$ to $sp^3$ enhances the spin-orbit by two orders of magnitude. Atomic
spin-orbit coupling in graphene is a very weak second order process, but H
adsorption can turn this into a first-order process. The spin-orbit coupling in
graphene becomes of the order of the atomic spin-orbit coupling
($\Delta^{at}_{SO}\approx$ 10 meV) and is comparable to that
of diamond.\cite{cas09}

\subsection{Vibrational properties}
\begin{figure}[htbp]
  \centering
  \includegraphics[width=8.0cm]{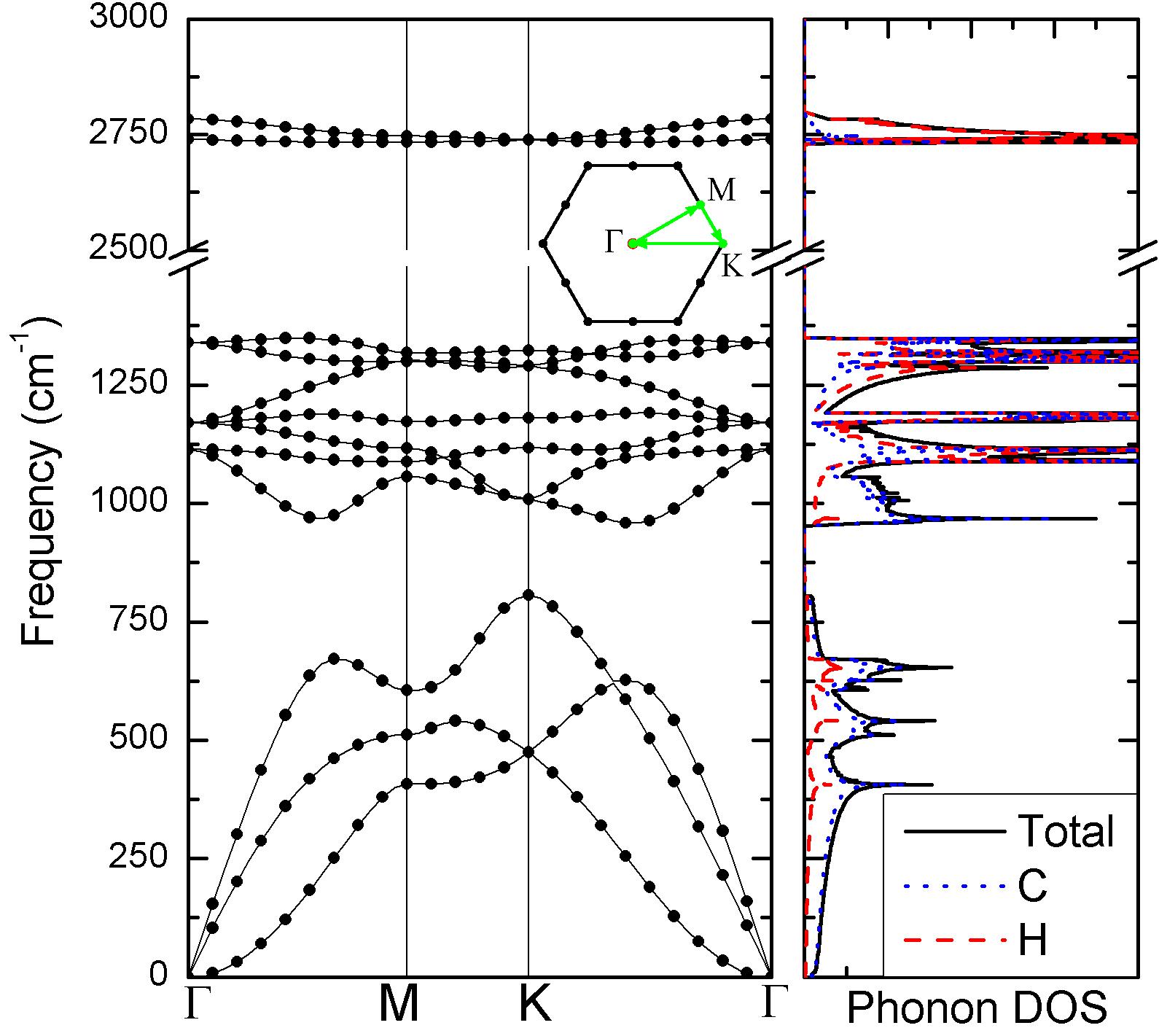}
  \caption{Phonon dispersion of graphane in the \textit{chair} configuration.
The dots are the directly calculated frequencies and the lines are interpolated
values. The inset shows the first Brillouin zone and the wavevector path used.
The right-hand side shows the phonon DOS.\label{fig-phon}}
\end{figure}

In Fig.\ \ref{fig-phon} the phonon spectrum of graphane in the \textit{chair}
configuration is shown. The absence of imaginary frequencies indicates that
this graphane isomer is stable. The phonon modes of graphane can be divided
into low, intermediate, and high-frequency groups of phonons.\cite{pee11} From
the projected phonon DOS, the high-frequency modes are identified as dominantly
H modes, as can be expected from the C–H stretching modes. The low-lying
(acoustic) modes are dominantly C modes while the intermediate-frequency group
of phonons have contributions from both types of atoms.

Savini et al. showed that upon hole doping, a giant Kohn anomaly
arises in the intermediate-frequency (optical) phonon spectrum of
graphane.\cite{sav10} This phenomenon was predicted to turn graphane into an
electron-phonon superconductor with a critical temperature above 90 K. This is a
consequence of the unique strength of the chemical bonds between the carbon
atoms and the large density of electronic states at the Fermi energy due to the
reduced dimensionality of graphane.

In addition the presence and characteristic properties of discrete breathers 
(DBs) which are spatially localized, large-amplitude vibrational modes in 
defect-free nonlinear lattices was investigated by Chechin et al.\cite{che14} It 
is found that  DB frequency decreases with increase in its amplitude, and it 
can take any value within the phonon gap and can even enter the low-frequency 
phonon band.

\subsection{Mechanical properties}
Graphene is the strongest material ever measured and it has also a remarkable
stiffness. Hydrogenation reduces the strength of graphene because the strong
aromatic bond network is replaced with single $\sigma$ bonds. The 2D Young's
modulus of graphane has been calculated\cite{lee10,top10,mun10} to be around
245 Nm$^{-1}$ which is substantially smaller than the
calculated\cite{lee10,top10} and experimental value\cite{lee08,koe10} for
graphene 340 Nm$^{-1}$. Experimental values for the Young's modulus of
graphane are not
available at present, but they are expected to be lower than the theoretical
value due to the presence of defects. This is comparable to the case of
fluorinated graphene where the theoretical estimate is about twice the
experimental value\cite{lee10,nai10}. Other isomers of graphane such as the
\textit{boat} and \textit{stirrup} configurations exhibit highly anisotropic
behavior.\cite{lee10,luc11} The Young's modulus can decrease to half its value
in particular directions and also the Poisson ratios are very
anisotropic.\cite{luc11}

Perfect graphane behaves elastically under strains up to at least 30\%, but
small defects can drastically reduce this limit\cite{top10}. For larger
strains, graphane shows irreversible changes and defects start to appear. The
out-of-plane stifness of graphane is also somewhat smaller than that of
graphene, leading to a higher surface roughness and thermal
contraction.\cite{nee11}

Thermal fluctuations of single layer graphane were investigated by Costamagna et al.~\cite{cos12} 
up to temperatures of at least 900 K. By analyzing the mean square value of the height fluctuations 
and the height-height correlation function for different system sizes and temperatures, they showed 
that graphane was an unrippled system in contrast to graphene where later follows the membrane theory. 
 This unexpected behavior persists  and is a consequence of the fact that in graphane the thermal energy 
 can be accommodated by in-plane bending modes, i.e., modes involving C-C-C bond angles in the buckled 
 carbon layer, instead of leading to significant out-of-plane fluctuations that occur in graphene.


\section{Synthesis}
Elias et al.~\cite{eli09} reported the  experimental fabrication of
graphane in 2009. Graphene samples were extracted using micromechanical
cleavage of graphite. These samples were either deposited on top of an oxidized
Si substrate or used as free-standing membranes for transmission electron
microscopy (TEM). The samples were first heated at high temperature to remove
all possible contamination. Cold hydrogen plasma at pressure 0.1 mbar
hydrogen-argon mixture (10\% H$_2$) was used for 2 hours to expose graphene
samples.

\begin{figure}
\includegraphics[width=8.5cm]{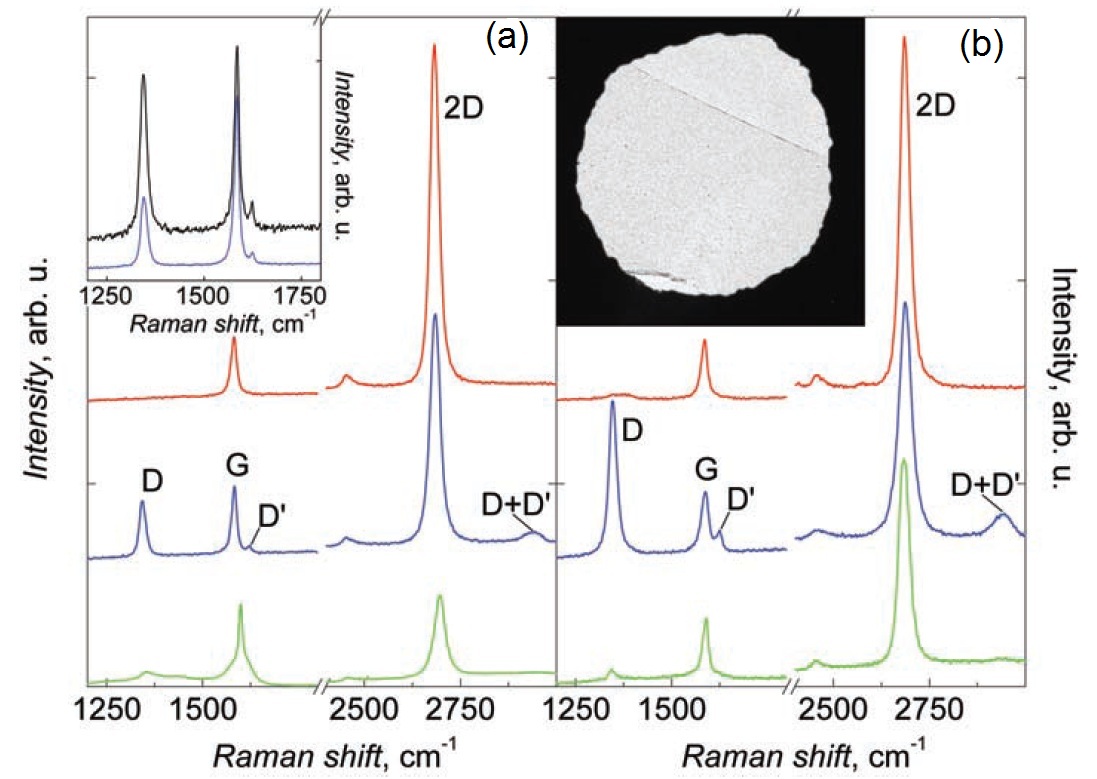}
\caption{\label{raman} (Color online) Changes in Raman spectra of graphene
caused by hydrogenation. The spectra are normalized to have a similar intensity
of the G peak. (a) Graphene on SiO$_{2}$. (b) Free-standing graphene. Red, blue,
and green curves (top to bottom) correspond to pristine, hydrogenated, and
annealed samples, respectively. Reproduced with permission from
ref. \onlinecite{eli09}. Copyright 2009 The American Association for the
Advancement of Science.}
\end{figure}

Raman spectroscopy is an effective tool to see the changes induced by
hydrogenated graphene. Fig.~\ref{raman} exhibits the evolution of Raman
spectra for graphene, hydrogenated  and annealed graphene samples.
Hydrogenation also results in sharp D (1342 cm$^{-1}$) and D$'$ ($\sim$1610
cm$^{-1}$) peaks and combination of both D+D$'$ peak around 2950 cm$^{-1}$  as
compared with disordered and nanostructured carbon-based materials. The D peak
in graphane appeared due to the breaking of the translational symmetry of
the C-C sp$^{2}$ bonds after the formation of the C-H sp$^{3}$ bonds. The  D
peak for both side hydrogenated graphene is twice larger than single sided
hydrogenated graphene, due to the formation of twice C-H bonds.

The metallic character of graphene from graphane can be recovered by annealing
graphane  in Ar atmosphere at 450 $^{\circ}$C for 24 hours as shown in
Fig.~\ref{raman}. Higher annealing temperature may damage the sample by
introducing structural disorder which broaden the intensity of the bands
indicating that the annealed graphene becomes hole-doped. All the defect
related peaks (D, D$'$, and D+D$'$) were strongly suppressed.

Wojtaszek et al.~\cite{woj11} created hydrogenated single and bilayer
graphene by an Ar-H$_2$ plasma produced in a reactive ion etching (RIE) system.
They
showed that the chosen plasma conditions can prevent damage to the graphene
sheet. The reported hydrogenation was 0.05\% which can be further improved.
The hydrogenation occurs due to the hydrogen ions from the plasma and not due
to the fragmentation of water adsorbates on the graphene surface by highly
accelerated plasma electrons.

Poh et al.~\cite{poh12} studied the production of partially hydrogenated
graphene using  thermal exfoliation of graphite oxide in H$_2$ atmosphere under
high pressure (60-150 bar) and temperature (200-500 $^{\circ}$C). They found
that a H$_2$ pressure of 100 bar at 500 $^{\circ}$C was the most efficient
reaction condition. This method does not require a plasma source (in contrast to
previous work~\cite{eli09}) and thus, a large amount of hydrogenated
graphene can be produced.

Guisinger et al.~\cite{gui09} used room temperature STM to investigate the
atomic hydrogen passivation of dangling bonds at the interface between single
layer graphene and silicon-terminated SiC which can strongly influence the
electronic properties of the graphene overlayer. Since hydrogen did not diffuse
through the graphene monolayer suggesting that the edge of graphene is
chemically bound to the reconstructed SiC(0001) surface.

Luo et al.~\cite{luo09} showed that the hydrogenation of graphene
layers by hydrogen plasma is reversible, even at its saturated hydrogen
coverage. The hydrogen coverage depends on various parameters such as
plasma power and process duration. The hydrogenation rate of graphene layers
is controlled by the hydrogenation energy barriers, which show a clear
dependence on the number of layers. It is  demonstrated, using Raman
spectroscopy, that the hydrogenation of bilayer and multilayer graphene is much
more feasible than that of single layer graphene on SiO$_2$/Si substrate. The
dehydrogenation also shows significant dependence on the numbers of graphene
layers and the amount of hydrogen coverage.

Wang et al.~\cite{wan10} reported  a new route to prepare high quality,
monolayer graphene by dehydrogenation of a graphane-like film grown by
plasma-enhanced chemical vapor deposition. The advantage of the plasma
deposition process is  its compatibility with wafer scale processing and
lithographical patterning as well as deposition on metal-coated silicon
substrates at temperature (at least 350 $^{\circ}$C) lower than that used
in thermal CVD processes. Using laser writing, graphane-like ribbons and squares
can be formed which is a step towards to design of an integrated circuit based
on all-carbon electronics.

Jones et al.~\cite{jon10} have synthesized partially hydrogenated
graphene on both sides and on single side of graphene by electron irradiation
of graphene having chemisorbed H$_2$O and NH$_3$ on the layer. Hydrogenation
was proposed due to H$^{+}$ ions and H radicals resulting from the fragmentation
of H$_2$O and NH$_3$ adsorbates by backscattered and secondary electrons.

Ryu et al.~\cite{ryu08} realized hydrogenation of graphene by
electron-induced dissociation of hydrogen silsesquioxane (HSQ). Hydrogenation
occurred  at a higher rate for single than for double layers due to the
enhanced chemical reactivity of a single sheet of graphene. The probability of
chemisorption of hydrogen atoms on single layers is at least 0.03 at room
temperature which is 15 times larger than for that for bilayers.

Theoretically, many authors proposed different approaches for the
hydrogenation graphene.

Using density functional theory, Zhou et al.~\cite{zho09,zh109} proposed
semi-hydrogenated graphene by applying an external electric field to a fully
hydrogenated graphene which can remove H atoms from one side of graphane.
Semi-hydrogenated graphene, also known as "graphone" is a  ferromagnetic
semiconductor with an indirect band gap of 0.43 eV. But it was shown that
free-standing graphone will roll up.\cite{nee13} In addition, Ao et
al.~\cite{ao10} showed that hydrogenation of graphene layers is easier
when an electric field that acts as a catalisor for dissociative adsorption
of H$_{2}$, is applied.

Leenaerts et al.~\cite{lee09} performed ab initio density-functional
theory calculations to investigate the process of hydrogenation of a bilayer of
graphene. 50\% hydrogen coverage is possible in case that the hydrogen atoms
are allowed to adsorb on both sides of the bilayer. In this case, the weak van
der Walls forces between the graphene layers are replaced by strong interlayer
chemical bonds. At maximum coverage, a bilayer of graphane is formed which has
similar electronic properties  to those of a single layer of graphane due to
the same hybridization.  Such a bilayer can be viewed as the thinnest
layer of diamond which was recently realized experimentally.\cite{raj13}
Samarakoon et al.~\cite{sam10} studied the electronic properties of
hydrogenated bilayer graphene by applying a perpendicular electric bias using
DFT calculations which leads  to a transition from semiconducting to metallic
state. Also, desorption of hydrogen from one layer in bilayer graphane
yields a ferromagnetic semiconductor with a tunable band gap.

Balog et al.~\cite{bal10} demonstrated that a bandgap can be opened in
graphene by inducing patterned hydrogen chemisorption onto the Moir\'{e}
superlattice positions of graphene grown on an Ir(111) substrate.
Combined STM and ARPES results demonstrated that the observed hydrogen adsorbate
structures were stable well above room temperature. The band gap was not very
sensitive to the exact hydrogen adsorbate structure.  These experimental
results also  support the concept of confinement-induced gap opening by
periodic lattice perturbations. The gap was induced at the Fermi energy and was
of sufficient size, such that it can be used for electronic applications at room
temperature. Balog et al.~\cite{bal09}  presented STM studies to
reveal the
local adsorbate structure of  atomic hydrogen on the basal plane of graphene
on a SiC substrate. Four types of configurations were formed by  hydrogen pairs
after exposing to a 1600 K D-atom beam. At low coverage the formation of
hydrogen dimers occurred, while at higher coverage random adsorption into larger
hydrogen clusters was observed. They also found that the tunneling current
of the STM tip
can induce hydrogen desorption, which implies that the hydrogen atom is weakly
bound
to the graphene surface. The hydrogenation was reversible
by annealing the substrate to 800 $^{\circ}$C. The adsorption of atomic
hydrogen onto the desired areas on the surface and to form nanopatterns via
tip-induced desorption of hydrogen, opens the possibility of electronic and
chemical functionalization of graphene surfaces via hydrogenation.
Haberer et al.~\cite{hab10} observed by angle-resolved photoemission
spectroscopy that a gap can be opened up to 1.0 eV for a hydrogen coverage of
8 percent of graphene on Au.


\section{Engineering the Properties of Graphane}

Controlling the electronic property of a material is essential in device
technology. Monolayer materials, with their reduced dimensionality, provide a
suitable playground for the modification of the characteristic properties of
materials at atomic scale. In order to control or modify the properties of
graphane, which is a nonmagnetic semiconductor, one can suggest
several techniques such formation of graphanes with different stoichiometry,
creating various vacancies, doping with foreign atoms, application of strain
and dimensional reduction.

\subsection{Graphanes with Various C/H Ratios}

If the adsorption of H atoms is only allowed at one side of the graphene layer,
some interesting theoretical materials can be formed. Hydrogenation of a single
sublattice gives rise to a ferromagnetic material that was called
graphone.\cite{zho09} The stability of graphone is very
weak\cite{pod11,puj11,nee13}
because the H atoms try to form pairs. Its practical use is therefore rather
questionable. In addition, some studies such as controlled H-domain formation 
on Ir(111) supported graphene\cite{bal13}, excitation-induced hydrogen 
dissociation from the graphane surface\cite{ban13} were already reported by 
experimental groups.

\begin{figure}[htbp]
  \centering
  \includegraphics[width=5.0cm]{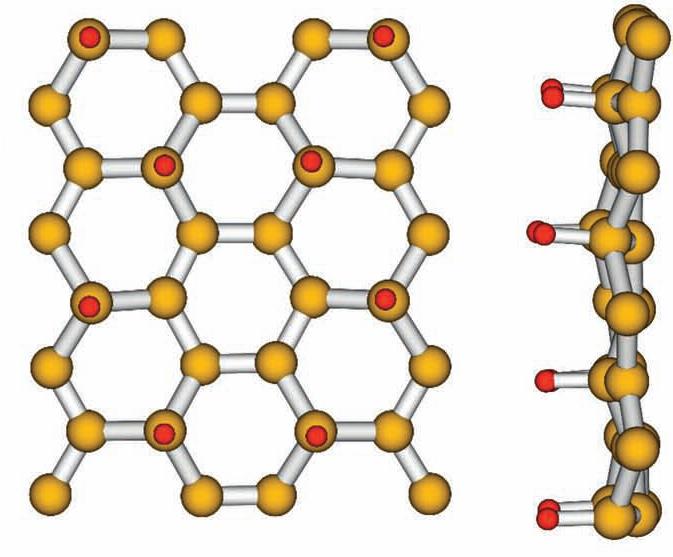}
  \caption{Top and side view of the C$_4$H crystal structure. From Ref. \
\onlinecite{hab11}.\label{fig-C4H}}
\end{figure}

A more promising case of single-side adsorption is C$_4$H. This structure is
built from H pairs in the very stable \textit{para} configuration in which
atoms are adsorbed on opposite C atoms in the graphene hexagons. When this
structure is repeated, a $2\times$2 superstructure arises which is very stable
(see Fig.\ \ref{fig-C4H}).\cite{hab11} In 2011, this 2D C$_4$H crystal was
synthesized by Haberer et al.\cite{hab11} on a Au substrate. Hydrogen
plasmas were used to hydrogenate the graphene sample to a H/C ratio of 1:4. DFT
calculations show that the resulting material contains an electronic band gap
of 3.5 eV which is similar to graphane.\cite{hab11,li12}

It is also possible to hydrogenate multilayer graphene. This can be done in two
different ways. First, the layers can all be completely hydrogenated to form
stacked sequences of e.g. \textit{chair} isomers, and second, only the top and
bottom layer are hydrogenated from one side, resulting in a thin diamond films.

In addition, Fokin et al. studied the effect of $\sigma$-$\sigma$
and $\pi$-$\pi$ interactions in the electronic properties of single layer and 
multilayer [n]graphanes at the dispersion-corrected density functional theory 
(DFT) level.\cite{fok11} It was found that graphanes show quantum confinement 
effects as the HOMO-LUMO gaps decrease from small to large graphane structures, 
asymptotically approaching 5.4 eV previously obtained for bulk graphane. 
Similarly, Alonso \textit{et al.} investigated the nature and origin of 
dispersion interactions for the benzene and cyclohexane dimers by using 
dispersion-corrected density functional theory, energy decomposition analysis, 
and the noncovalent interaction (NCI) method. Their NCI analysis revealed 
that the dispersion interactions between the hydrogen atoms are responsible for 
the surprisingly strong aliphatic interactions.


Completely hydrogenated graphene layers can be stacked to form multilayered
systems or 3D crystals. The H atoms of the different layers have a repulsive
effect which causes the layers to align in such a way that the H atoms of one
layer do not overlap with the H atoms of the neighboring graphane layer. The
most stable multilayered graphane system is built from AA-stacked
\textit{chair} isomers.\cite{roh11,wen11} However, under pressure, other
graphane
configurations can become more favorable as building blocks for 3D graphane
crystals. Wen et al. calculated that for pressures above 10 GPa, the
\textit{stirrup} isomer becomes more stable due to a more efficient interlayer
H arrangement.\cite{wen11} Pressure was also shown to increase the electronic
band gap up to 20 GPa.\cite{wen11}\\

\begin{figure}[htbp]
  \centering
  \includegraphics[width=7.0 cm]{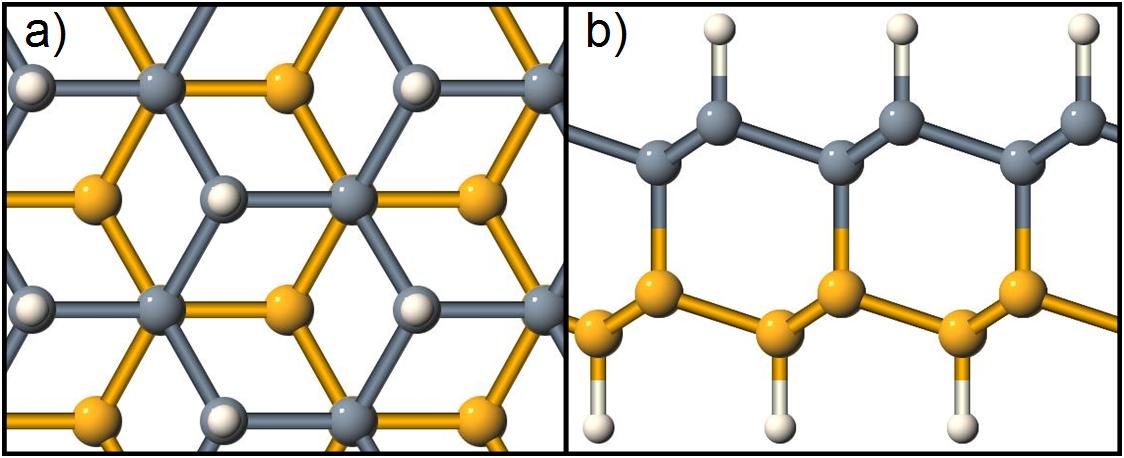}
  \caption{Top and side view of bilayer diamane structure. From Ref. \
\onlinecite{lee09}
  \label{fig_biGa}}
\end{figure}

When the hydrogenation process of few-layer graphene (FLG) is restricted in
such a
way that only the outer layers are exposed to H atoms, it is possible to turn
FLG in thin diamond films. Such films have been dubbed 'diamane'. Hydrogen
adsorption on the top layer changes the hybridization of the interacting C
atoms to $sp^3$, but also the neighboring C atoms will undergo a substantial
hybridization change.\cite{lee09} When the H coverage is large enough, it
becomes favorable for the non-hydrogenated C atoms in the outer layers to form
covalent bonds with the underlying layers (see
Fig.~\ref{fig_biGa}).\cite{lee09} The formation of interlayer C-C bonds is
easiest achieved in AA or ABC-stacked graphene layers\cite{zhu11,odk13,kva14},
but has also been theoretically predicted for twisted graphene
bilayers.\cite{mun12} In addition, Mapasha et al. investigated hydrogenation of 
bilayer graphene using GGA-PBE functional and four variants of non-local van der 
Waals density functionals vdW-DF, vdW-DF2, vdW-DF-C09x, and 
vdW-DF2-C09x.\cite{map13}

For ordered FLG slabs the process can be continued until all C atoms in the
system have $sp^3$ hybridization. In this ways the hydrogenation process can
turn FLG into a thin diamond film,\cite{zhu11,odk13,kva14} as has been
experimentally confirmed.\cite{raj13} The band gap of diamane decreases with
the number of layers\cite{che11,zhu11}
and does not converge to the band gap of bulk diamond. The reason for this is
that the size of the band gap is determined by the position of the CBM which
corresponds to a surface state (see Fig.\ \ref{fig_WF}). As predicted by
Samarakoon et al., the band gap of bilayer diamane can also be tuned by
the application of an
electric field. The gap decreases monotonically with increasing field strength
and a semiconductor to metal transition is observed at 1.05
V/$\text{\AA}$.\cite{sam10}

\subsection{Vacancy Formation}

As it was shown before even the strong covalent bonds of graphene can be
broken and various vacancies are formed due to e.g. continuous exposure
to the
high-energy electron beam\cite{kot11} or irradiation with low-energy
Ar$^{+}$.\cite{ahl13} Moreover, the fabrication of large graphene sheets
having a high-density of nanoscale holes or multiple carbon vacancies is another
landmark in controlling the electronic properties of
graphene.\cite{bai10,lah10} Therefore, motivated by these recent
advances, controllable modification of the electronic properties of graphane,
that has weaker covalent bonds as compared to graphene, can be utilized for
different applications.

\begin{figure}[]
  \centering
  \includegraphics[width=8.5cm]{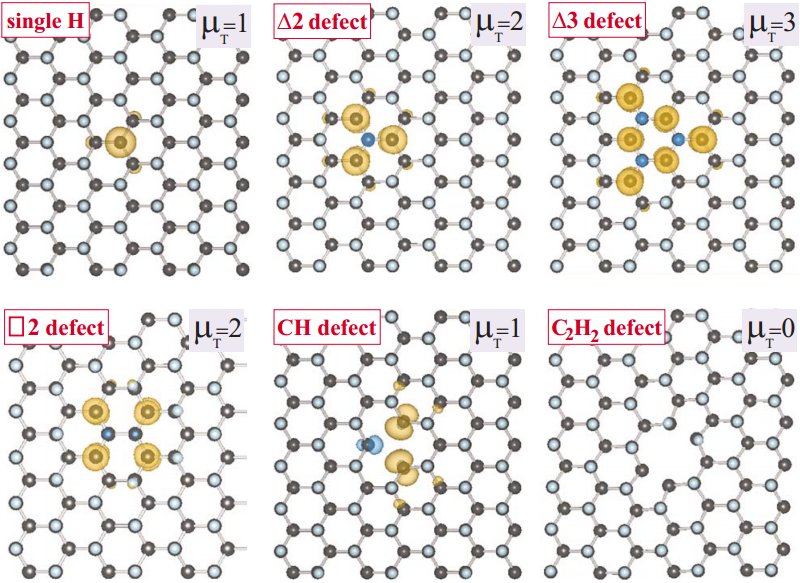}
  \caption{Magnetic properties of graphane with various defects created by
removing H atoms and C-H pairs (from Ref. \
\onlinecite{sah10}).\label{fig-defect}}
\end{figure}

As was shown by Sahin et al.\cite{sah09}, desorption of a
single H atom from the surface of graphane requires 4.79 eV energy per
atom. Formation of each H-vacancy leaves a half-filled sp$^{3}$-like orbital
on the graphane surface and therefore the bandgap is reduced by the defect
states
that appear around the Fermi level. Here it is worth to note that each single
H-vacancy creates a local magnetic moment of 1 $\mu_{B}$. As shown in Fig.
\ref{fig-defect}, one can obtain large spin polarizations by creating
triangular-shaped H-free regions on graphane. In addition, as long as the
number of freed H atoms (C-H pairs) from A and B sublattice is the same there
is no net magnetic moment in the ground state of the structure. Ferromagnetic
domains with large net magnetic moments on graphane are useful for future data
storage and spintronics applications. Likewise, Moaied et
al.\cite{moa14} reported that hydrogenation of
single and multilayer graphene surfaces may result in a ferromagnetic state. It
was also predicted that the Curie temperature strongly depends on the size of
the hydrogenated region. Similarly, a recent study
by Walter et al. showed that created vacancy domains on
fluorographene, which is the fluorinated counterpart of graphane, possess a band
gap which was reduced by midgap states resulting in a blue light emitting
ground state which is not the case for defect-free material.\cite{and14}

Creation of one-dimensional graphene/graphane superlattices with a tunable
electronic bandgap can also be realized by selective desorption of H
atoms. Hernandez-Nieves et al.\cite{her10} predicted, using ab initio
calculations, that one-dimensional H-free regions serve as freestanding
graphene nanoribbons, and their magnetic ground state can be tuned via
particular arrangements of the H atoms at the edges. In recent work of Ray
et al.,  spintronics applications of
the graphene supported graphone/graphane bilayer structure was investigated
experimentally. It was shown that hydrogen plasma treatment of vertically
aligned few layer graphene results in the formation of graphane/graphone layers
with ferromagnetic ground state.\cite{ray14} Therefore graphone/graphane
bilayers with their hydrogen concentration dependent magnetic properties are
promising for nanoscale spintronic device applications. It appears that
the formation of H vacancies on graphane not only suitable for providing spin
polarization but can also be used for engineering the electronic
characteristics.

\subsection{Doping by Foreign Atoms}

Although impurities such as single atoms, molecules and small clusters on
ultra-thin materials appear as undesirable residua from the synthesis process,
these impurities can also be utilized for the modification of
electronic and magnetic properties of these materials.

Lu et al. showed that the electronic structure of fluorine doped
graphane is very sensitive to the doping configuration, due to the competition
between anti-bonding states and nearly surface states.\cite{lu09} An
interesting application of graphane, as a PNP transistor, was proposed
by Gharekhanlou et al.\cite{gha10} . Using graphane with hydrogen
deficiency to reduce the band gap, it was shown that, within the approximation
of the Shockley law for junctions, an exponential ideal I-V characteristic is
expected and the curvature of the collector current characteristics shows good
agreement with an ideal bipolar transistor.

Moreover, Hussain et al. studied the stability and possibility of
hydrogen storage applications of Ca-doped graphane. They found that Ca-doped
graphane structure remains stable even at high temperatures and hydrogen
storage capacity of a monolayer Ca-doped graphane with a doping concentration
of 11 \% of Ca on a graphane sheet, and a reasonably good H$_{2}$ storage
capacity of 6wt \% could be attained.\cite{hus12} In addition the same
group
reported that neither a pristine graphane sheet nor the sheet defected by
removing a few surface H atoms have sufficient affinity for either H$_{2}$S or
NH$_{3}$ gas molecules. However, a graphane sheet doped with Li adatoms shows a
strong sensing affinity for both mentioned gas molecules.\cite{hus14}

Doping of graphane not only results in the modification of the electronic
bandgap, but also can yield the emergence of novel phases in the material.
Savini et al. predicted by using first-principles calculations that
p-doped graphane is an electron-phonon superconductor and the presence of a
giant Kohn anomaly in the optical phonon dispersions, that originates from a
large density of electronic states at the Fermi energy and strong carbon-carbon
bonds, results in a superconducting phase.
\cite{sav10} In addition, using mean-field theory, Loktev et
al.\cite{lok11} considered
superconducting properties of multilayer and single layer graphane by taking
into account fluctuations of the order parameter. They showed that, even for
low doping cases, in the single-layer and multilayer case the critical
temperature is predicted to be 100 K and 150 K, respectively.

\begin{figure}[]
  \centering
  \includegraphics[width=8.5cm]{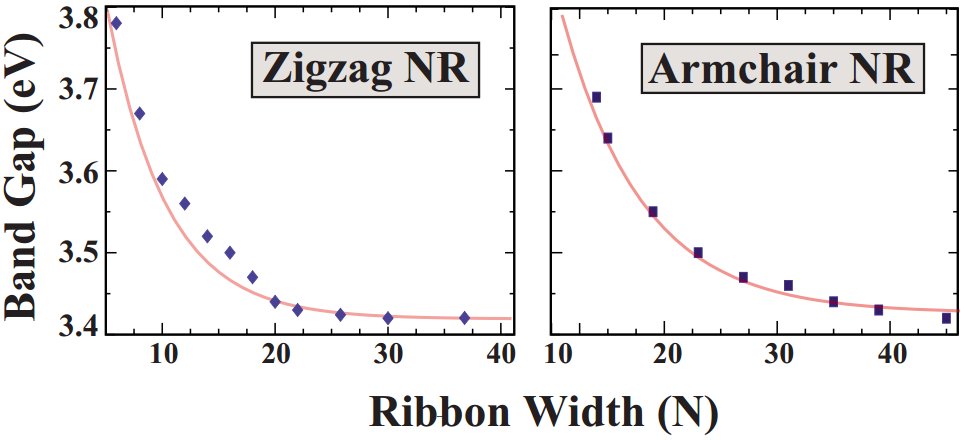}\\
  \includegraphics[width=8.5cm]{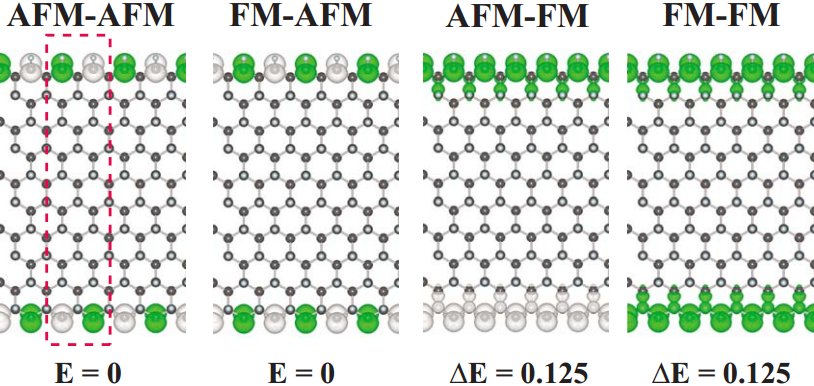}
  \caption{(Upper row) Energy
bandgap of zigzag and armchair nanoribbons as a
function of ribbon width. (Lower row) Magnetic ground state analysis of zigzag
graphane nanoribbons from Ref.\ \onlinecite{top10}.\label{fig-ribbon}}
\end{figure}

Da et al.~\cite{dah11} studied the magnetic properties of graphane
by substituting the C atoms along with its attached H atoms with transition
metal (TM) atoms.  It was also shown that a maximum magnetic moment of 3.5
$\mu_{B}$ can be reached by embedding Mn in graphane. In another DFT 
study by AlZahrani revealed that the hollow-site adsorption configuration on 
graphane is the most preferable one for Mn atoms.\cite{alz12} Also, a
heterojunction structure embedded with nickel and vanadium within graphane
suppress the spin-down current while the spin-up current appears at the
negative bias voltage, resulting in a spin current diode. Yang~\cite{yan10}
showed that creating a vacancy of a hydrogen atom generates a magnetic moment of
1 $\mu_{B}$ and if this vacancy is occupied by transition metal elements, the
magnetic moment can enhanced many fold. Leenaerts et al.~\cite{lee13} investigated the
 effect of substitutional doping of fluorographene with boron and nitrogen atoms on its electronic and
magnetic properties using first-principles calculations. Boron dopants acted as shallow acceptors and caused 
hole doping but no changes in the magnetic properties were observed. Nitrogen dopants acted 
as deep donors and gave rise to a magnetic moment, but the resulting system became chemically unstable. 
These results are opposite to what was found for substitutional doping of graphane in which case B substituents induce
magnetism and N dopants do not.

In addition, bandgap excitations graphane and its derivatives was studied by 
Nelson \textit{et al.}\cite{nel13} The nonradiative lifetime and radiative line 
width of the lowest energy singlet excitations in pure and oxidized graphanes 
were determined by performing nonadiabatic molecular dynamics combined with 
time-dependent density functional theory.

\subsection{Ribbons}

Advances in experimental techniques have also made the synthesis of
ultra-narrow one dimensional structures "few-atoms-wide ribbons" possible.
Dimensional reduction of graphane is another efficient way of tuning its
electronic properties. As reported by Sahin et al.\cite{sah10}
nanoribbons (NRs) of graphane have some distinctive properties. It
was shown that H-terminated armchair and zigzag graphane NRs display a band-gap
reduction with increasing width (see Fig. \ref{fig-ribbon}). This behavior
could be
fitted to an expression $E_{gap} = 3.42 + \alpha exp(-N\beta)$ eV. Here
$\alpha$ and $\beta$ are fitting parameters and
are found to be 1.18 (2.15) and 0.19 (0.14) for zigzag (armchair) graphane NRs.
 While bare zigzag graphane NR is an antiferromagnetic semiconductor with an
indirect band gap relatively smaller than that of 2D graphane, armchair NRs are
nonmagnetic semiconductors. However, upon the saturation of edge atoms both NR
types become direct band-gap semiconductor having a nonmagnetic ground state.

Moreover, Zhang et al.\cite{zha12}, using the density functional
theory methodology,  predicted that the band gap of graphane nanoribbons can be
tuned linearly with strain regardless of their widths or edge structures.
Moreover, the band gap of the graphane nanoribbon is more sensitive to
compressive than tensile deformation, which mainly originates from the shift of
its valence band edge under strain. In addition, another interesting property,
the presence of two types of nearly-free-electron states in graphane NRs was
predicted by Liu et al.\cite{liu11} It was also shown by Wu et
al.\cite{wu14} that by forming one-dimensional H-vacancy chains in graphane
one can create various patterns of graphane/graphene NRs with tunable
bandgaps.\cite{her10}

\subsection{Strain Engineering}

Due to the  atomic scale thickness of ultra-thin materials characteristic
properties of them are quite sensitive to lattice deformations. As an efficient
way, "strain engineering" of two-dimensional materials has been widely
studied to tailor the electronic properties of materials and improve their
optical properties.

\begin{figure}[]
  \centering
  \includegraphics[width=8cm]{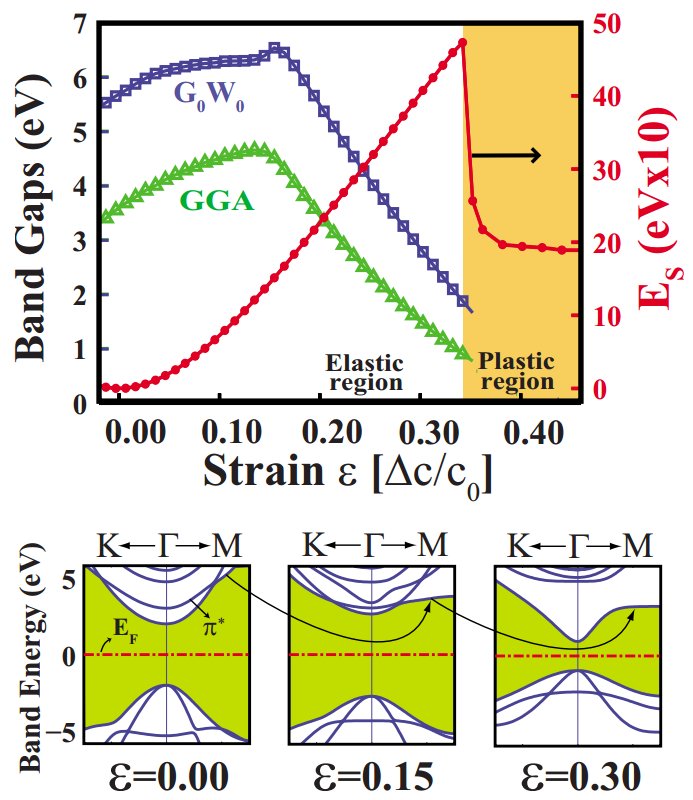}
  \caption{Evolution of the energy bandgap (upper) and electronic dispersion
(lower) as a function of strain (from Ref. \
\onlinecite{sah10}).\label{fig-strain}}
\end{figure}

Elastic properties of graphane were first studied by Topsakal
et al.\cite{top10}, using DFT-based strain energy calculations in the
harmonic elastic deformation range. It was found that graphane is a quite
stiff material and although the applied strain has negligible effect on strong
C-H bonds, stretched C-C bonds yield a dramatical modulation in the conduction
band states (see Fig.~\ref{fig-strain}). In addition, quantum-mechanochemical
reaction-coordinate simulations were performed by Popova et
al.\cite{pop11} to investigate the mechanical properties of graphane. They
showed that hydrogenation of graphene drastically influences the behavior and
numerical characteristics of the body, making tricotage-like pattern of
graphene failure less pronounced and inverting it from the zigzag to armchair
mode as well as providing less mechanical resistance of graphane in total.

Recently, Peng et al.\cite{pen13} studied the effect of dispersion
forces on the mechanical properties of graphane. By performing DFT calculations
with DFT-D2 approximated van der Waals (vdW) corrections they showed that vdW
has little effect on the geometry ($<$0.4\%) and the mechanical
properties, including ultimate stresses (2\%), ultimate strains (8.7\%),
in-plane stiffness (1\%) and Poisson ratio (3\%).

\section{Possible Applications}

As the  demand for energy is increasing across the world and concerns over
CO$_2$ emission to climate change, hydrogen has emerged as a potential
candidate as energy carrier due to its combustion waste product
water.~\cite{sch01} The advantage of using graphane as H$_2$ storage material
is its nano-size, large stability and relatively stronger graphane-metal
binding. Recently, Hussain et al.~\cite{hus14} showed that defected
graphane functionalized both sides by poly-lithiated species CLi$_3$ and
CLi$_4$, a storage capacity of more than 13 wt \% can be achieved. Furthermore,
the binding energies of the absorbed H$_2$ are in the range 0.25 eV $-$ 0.35
$eV$, which is well suited for an efficient H$_2$ storage. Hussain et
al.~\cite{hus12} also  studied the effect of strain on the adsorption of
hydrogen on the lithium doped graphane system. For strain-free case, at most 3
hydrogen molecules can be adsorbed on each Li resulting in a storage capacity
of 9.37 wt\%. While applying biaxial asymmetric strain, a stronger binding
between Li-graphane occurs. Now at most 4 hydrogen molecules can be adsorbed on
each Li enhancing the storage capacity to 12 wt\% beyond the Department of
Energy target of 2017. In addition, Xiao et al.\cite{xia14} examined CO$_{2}$ 
adsorption over various N-substituted/grafted graphanes to identify the 
promotional effects of various N-functionalities have on the adsorption 
characteristics using DFT. It was found that the presence of co-adsorbed 
H$_{2}$O on the surface promotes CO$_{2}$ adsorption on both N- and 
NH$_{2}$-sites, with highly exothermic adsorption energies.

Frictionless, lubricant or near-frictionless surfaces may find 
applications in the various fields of materials science and engineering.
Using the Prandtl-Tomlinson model together with ab initio calculations 
Cahangirov \textit{et al.} investigated the interlayer interaction under 
constant loading force and derived the critical stiffness required to avoid 
stick-slip behavior.\cite{cah12} It was also shown that graphane and similar 
layered structures have low critical stiffness even under high loading forces 
due to their charged surfaces repelling each other. Similarly, Wang \textit{et 
al.} calculated the friction characteristics of graphane and fluorographene. 
Their calculations revealed the general mechanism of atomic-scale friction in 
various graphene-based layered materials where the interlayer interaction is 
dominated by van der Waals and electrostatic interactions.\cite{wan13}


Kim et al.\cite{kim14} studied monolayer halogenated graphane
using first-principles calculations. Different configurations of hydrogen and
halogen atoms (F, Cl, Br) attached to graphene break the inversion symmetry and
hence give piezoelectricity. (C$_2$HF)$_n$ polar conformation was most
energetically stable with piezoelectricity due to the change of the electron
distribution around the F atoms. (C$_2$HCl)$_n$ showed an enhancement in the
piezoelectricity but the stability was degraded.

Thermoelectric materials, that have gained importance in recent years, can be
used for the development of new cooling and power generation methods. Using
thermoelectric materials one can convert the thermal difference into
electrical energy or use electrical current for the creation of temperature
difference. For an efficient thermoelectric device good electrical conductivity
and low thermal conductivity is essential. Considering these requirements Ni
et al. \cite{ni09} investigated the possible use of graphane as a
thermoelectric device and predicted that disordered armchair
graphane NRs are promising candidates for constructing thermoelectric
materials. The high figure of merit, low cost, and easy synthesis of graphane
make it a viable choice for thermoelectric applications.


The detection of explosives is one of the main concern for a secure and safe
society. Trinitrotoluene (TNT) which is the main component in explosives, is
toxic to the environment and a possible source of cancer in humans. Recently,
electrochemical methods have emerged due to its rapid response time,
portability, accuracy and lower cost than other analytical methods. Seah
et al.~\cite{sea14} used partially hydrogenated graphene and graphene as
a sensing platform for TNT in seawater. They found graphene is more sensitive
due to large aromatic rings which favor surface accumulation or larger
preconcentration of analytes. Moreover, Tan et al.~\cite{tan13} have
demonstrated, using electrochemical methods, that graphane exhibits different
electrochemical behavior towards oxidation/reduction of important biomarkers,
such as ascorbic acid, dopamine and uric acid when compared to ordinary
graphene.

\section{Conclusions}

In this overview we have discussed the so-called graphane from synthesis to
applications. We have showed that one-by-one hydrogenation of graphene results
in graphane which is a stable, direct bandgap insulator with strong C-H bonds.
We also showed that for the synthesis of graphanes one can choose various
methods such as hydrogen plasma exposure of graphene, thermal exfoliation of
graphene oxides, STM-assisted hydrogenation of graphene, plasma-enhanced
CVD and electron-induced dissociation of HSQ on graphene. We have also
presented that electronic and magnetic properties of graphene can be modulated
by various techniques such as synthesis of graphanes with different C/H ratios,
formation of vacancies, application of strain and dimensional reduction.
Finally, we showed that hydrogenated graphenes may also found applications in
various fields such as hydrogen storage, piezzo-electricity,
thermo-electrics, explosive detection and biosensing devices. We believe that
future studies on
functionalized graphane will reveal many interesting properties of this
material.

\begin{acknowledgments}

This work was supported by the Flemish Science Foundation (FWO-Vl) and
the Methusalem foundation of the Flemish government. H.S. is supported by a
FWO Pegasus Long Marie Curie Fellowship.

\end{acknowledgments}

\end{document}